# A study linking patient EHR data to external death data at Stanford Medicine

Alvaro Andres Alvarez Peralta, Priya Desai, Somalee Datta

Research IT, Technology & Digital Solutions, Stanford School of Medicine and Stanford Health Care

## Abstract:

This manuscript explores linking real-world patient data with external death data in the context of research Clinical Data Warehouses (r-CDWs). We specifically present the linking of Electronic Health Records (EHR) data for Stanford Health Care (SHC) patients and data from the Social Security Administration (SSA) Limited Access Death Master File (LADMF) made available by the US Department of Commerce's National Technical Information Service (NTIS).

The data analysis framework presented in this manuscript extends prior approaches and is generalizable to linking any two cross-organizational real-world patient data sources. Electronic Health Record (EHR) data and NTIS LADMF are heavily used resources at other medical centers and we expect that the methods and learnings presented here will be valuable to others. Our findings suggest that strong linkages are incomplete and weak linkages are noisy i.e., there is no good linkage rule that provides coverage and accuracy. Furthermore, the best linkage rule for any two datasets is different from the best linkage rule for two other datasets i.e., there is no generalization of linkage rules. Finally, LADMF, a commonly used external death data resource for r-CDWs, has a significant gap in death data making it necessary for r-CDWs to seek out more than one external death data source. We anticipate that presentation of multiple linkages will make it hard to present the linkage outcome to the end user.

This manuscript is a resource in support of Stanford Medicine STARR (STAnford medicine Research data Repository) r-CDWs. The data are stored and analyzed as PHI in our HIPAA-compliant data center and are used under research and development (R&D) activities of STARR IRB.

## Background:

We will present a deterministic linking of Stanford hospital patient data with an external death data source with the purpose of using the linked data in research Clinical Data Warehouses (r-CDWs). Neither data source was originally designed to meet academic medical center research needs. The first data source, EHR, is designed for patient care. The second data source, SSA death data, was designed to prevent fraud and abuse of federally funded benefits provided by SSA and other federal agencies.

Our methodology is based on two requirements. Firstly, the underlying software framework needs to support linking two cross-organizational data and, therefore, must be flexible and extensible. Secondly, the quality of linking should be computable and shareable with end-users of r-CDWs in the spirit of good data stewardship. In our attempt to develop such a methodology, we review several papers in patient linking [Just2016, Ruppert2016, Riplinger2020,

Grannis2022], pick ones that are suitable to datasets like ours [Kho2015, Just2016, Maaroufi2018], build on data cleaning/normalization approaches [Kho2015, Maaroufi2018] and extend patient linking data analysis frameworks [Sequoia2018].

While the methodology presented in this manuscript is broadly generalizable, we will present the analysis on linking EHR data from Stanford Health Care Epic Clarity to the Social Security Administration (SSA) Limited Access Death Master File (LADMF) made available by the US Department of Commerce's National Technical Information Service (NTIS). The purpose of the linked data is to support Stanford Medicine STARR (STAnford medicine Research data Repository, https://starr.stanford.edu/) r-CDWs. The data are stored and analyzed as PHI in our HIPAA-compliant data center and are used under research and development (R&D) activities of STARR IRB.

## About STARR:

Stanford Medicine STARR (STAnford medicine Research data Repository) contains electronic health record (EHR) data from the two hospitals, the Adult Hospital and Children's Hospital, and their affiliates, University Healthcare Alliance and Lucile Packard Healthcare Alliance. The two hospitals use Epic for patient care and share research-approved Epic Clarity data with the Stanford School of Medicine (SoM). These data are moved daily from the hospitals to STARR and processed for downstream consumption in research Clinical Data Warehouses (r-CDWs) e.g. the in-house data model formerly known as STRIDE [Lowe2009], OMOP [Datta2020], PEDSnet and contain data from 3-4 million patients.

One of the primary goals of STARR r-CDWs is to make the data fit for the widest variety of research. Much of the effort goes into ETLs that take raw EHR data to create analysis-ready data assets and support suitable cohort analysis tools *e.g.,* STARR Tools, OHDSI ATLAS Cohort Tool. Our in-house data model integrates patient data with NTIS LADMF. In Feb 2022, of ~4.56M patient records in the STRIDE database, ~200,000 are marked deceased, of these ~53,000 are from EHR data, and ~156,000 are from NTIS LADMF.

## Relevance of death data in research:

Electronic health records (EHR) data captures in-hospital death accurately, but often does not capture patient death when death occurs outside the hospital. Death as an outcome is an important determinant to understand the true burden of diseases like COVID, maternal health, and in the development of clinical decision-support algorithms. For example, in 2013, the Centers for Medicare and Medicaid Services (CMS) Hospital Readmission Reduction Program [HRRP2012] began to financially penalize US hospitals with excessive 30-day readmission rates, with the goal of improving patient care. This decision resulted in a large number of models to predict readmission rates. However, in a critical appraisal of 81 readmission models [Liu2021], authors found that 13 models (16%) did not account for mortality, 40 (49%) accounted for in-hospital mortality only, 5 (6%) accounted for post-discharge mortality only and 21 (26%) accounted for both.

## About SSA death data:

SSA compiles files of death information from many sources [CRSReport2021], including close relatives and representative payees, funeral homes, financial institutions, postal authorities, States, and other Federal agencies (e.g., Department of Veterans Affairs, Centers for Medicare and Medicaid Services) and added about 2.9 million new death reports to its records in 2019.

About ~78% of these death reports came from the states through a system called Electronic Death Registration (EDR) system. Using EDR, state vital records agencies first verify the decedent's name and Social Security number (SSN) against SSA's records and then submit a verified electronic death record that is automatically posted to SSA's Master Files of SSN Holders and SSN Applications, also known as the Numident. Death reports through EDR are timely (95% within 30 days) and highly accurate and are used across SSA's systems to automatically stop Social Security and SSI benefit payments to deceased individuals. Unfortunately, the NTIS DMF does not contain state death records and only contains only those death records obtained from non-state sources (i.e., obtained from close relatives, representative payees, other federal agencies, other state agencies such as state welfare offices, friends, and neighbors).

The full death data file is only accessible to certain federal benefit-paying agencies such as the Department of Defense, Department of Veterans Affairs, Department of Agriculture, Department of Housing and Urban Development, Centers for Medicare and Medicaid Services, Internal Revenue Service, and starting 2020, to National Institutes of Health for research and statistical purposes. NTIS sells a public version of the DMF, which, in addition to excluding death data from state vital statistics bureaus, also excludes records for individuals who died within the last three years. The LADMF is thus somewhat larger than the public DMF but still considerably smaller than SSA's full file of death information. NTIS further states that DMF does have inaccuracies and SSA does not guarantee the accuracy of the DMF file. Firstly, the SSA does not have a death record for all deceased persons. Secondly, in rare instances, it is possible for the records of a person who is not deceased to be included erroneously in the DMF.

Prior research studies [Levin2019, Navar2019] have also noted that the quality of LADMF changed substantially after 2011. Here is a summary of changes to SSA death data [Rothwell2018] explaining the change:
- SSA created the public DMF in 1980 in response to a 1978 Freedom of Information Act lawsuit [FOIA1978].
- In 2002, SSA began disclosing state EDR records on the public DMF.
- In 2010/2011, SSA began examining their disclosure of protected state records. Section 205(r) of the Social Security Act [CRSReport2021] prohibits SSA from disclosing State death records.
- Beginning Nov. 1, 2011, SSA decided it would withdraw state death records from the public file.

Aside from the data gaps, LADMF regulations place significant restrictions on the use and disclosure of the date of death (including individual data elements, such as a month, day, and year of death) of individuals during the three-calendar-year period beginning on the date of the individual's death. The fact of death is not regulated. If there is independent death information from a hospital source, confirming the date of death, the individual's data is not regulated under LADMF. The LADMF regulations are independent of HIPAA regulations and persist even in fully de-identified data. Inappropriate use is subject to penalties under federal regulations [15CFR1110.200]. These regulations make the dataset more complex for researchers to use in an academic research environment.

## Other sources of death data:

Other death data sources may be relevant for integration with STARR but they all pose their own sets of challenges for research use. One example is the California Department of Public Health (CDPH) Public Use Death Data, provided by the Vital Statistics Bureau (VSB). While the majority of Stanford's patients are from California, not everyone is. Census 2020 data shows that California has experienced its first significant decline in population and the Bay Area is one of the most impacted areas [CalMatters2022, CalPolicyLabBrief2022].

Another example is Datavant Mortality Data (https://datavant.com/resources/whitepapers/mortality-data-in-healthcare-analytics/) which combines obituary-mined data (algorithmic) with DMF. Validating algorithmically generated datasets can take significant work, more so when algorithms are proprietary.

For individual researchers, the highest quality death data can be obtained from National Center for Health Statistics (NCHS) National Death Index (NDI). The NDI is a NCHS centralized database of all U.S. deaths beginning in 1979 that is used to match a NCHS survey record to a NDI record of death according to NDI matching criteria. The NDI contains the following identifying information for each death, which are used for matching purposes: SSN, Name, DoB, Sex, Father's surname, State of birth, Race, State of residence, and Marital status. The NDI system returns to the researcher the date of death, state of death, and death certificate number. It can also return the cause of death, in ICD format, if the researcher has purchased the NDI Plus service. The NDI data completion rate is very high (https://www.cdc.gov/nchs/ndi/completion_status.htm).  The cost of using NDI is also high. NIH reimburses the costs to NIH-supported investigators if the research aims require linking [NOT-OD-20-057]. Access to NDI can not be requested at an organizational level which precludes using NDI for STARR r-CDWs.

NIH pragmatic trials collaboratory resource on clinical trials [Eisenstein4] presents an in-depth discussion on using external death data resources and provides a comparison of data elements from four sources DMF, NDI, Medicare Master Beneficiary File, California's non-comprehensive death file, and National Association for Public Health Statistics and Information Systems (NAPHSIS) Electronic Verification of Vital Events (EVVE) fact of Death (FOD).

## Data linkage methods:

There are essentially two types of patient matching/linking algorithms, deterministic [Kho2015, Just2016, Maaroufi2018] and probabilistic [Sayers2016, HaggerJohnson2017, Bannay2021]. Within a deterministic matching category, basic matching demands that the underlying data elements must match exactly or have an exact partial match (e.g., the initial of the Middle Name or the last four digits of SSN). A more sophisticated deterministic matching approach can support alphabet or digit transpositions, and typographical errors. Then there are probabilistic matching approaches that estimate match probability based on the degree of similarity between records. For example, if we were matching a claims dataset and a EHR dataset or two EHR datasets, both datasets will include diagnosis, procedures, and medications. In our case, we have very little overlap between the information content of the two datasets. Therefore, for the linking, we will focus on deterministic approaches.

STARR Tools, previously known as STRIDE [Lowe2009], links patient data from two Stanford hospitals (first linked using MRN), and NTIS LADMF using two (unpublished) rules:
- SSN + DoB
- SSN + First Name + Last Name + (YYYY of) DoB

Of the ~200,000 marked deceased in the in-house database [Lowe2009], only ~4000 patients are marked deceased in both EHR and NTIS LADMF data sources.

Kho et al [Kho2015] linked (EHR) data across multiple sites in a large metropolitan area in the United States (Chicago, IL) using four linkage rules from Name, SSN, and DoB:
- First Name + Last Name + DoB
- DoB + SSN
- Last Name + SSN
- (first 3 characters of) First Name + (first 3 characters of) Last Name + (soundex) First Name + (soundex) last name + DoB + SSN

This was also the first real-world application of privacy-preserving linking (PPRL). While Kho et al used only four rules, the underlying software application, Distributed Common Identity for the Integration of Regional Health Data – DCIFIRHD, creates hashes using 17 rules.

Maaroufi *et al* [Maaroufi2018] present a single rule for the French Rare Diseases Registry using Name, DoB, and Gender:
- First Name + Last Name + DoB + Gender

Levin et al [Levin2019] linked NTIS LADMF with New Jersey Department of Health Office of Vital Statistics and Registry data used three linkage rules to using Name, SSN and DoB:
- SSN
- (first 3 characters of) First Name + (first 3 characters of) Last Name + DoB
- (soundex) First Name + (soundex) last name + DoB

Privacy-preserving linking tools are now available from multiple providers [Aronson2020]. Two of the largest consortiums that Stanford participates in, PCORNet (via PEDSNet) and N3C (https://covid.cd2h.org/PPRL), have adopted Datavant commercial application that generates hashes similar to DCIFIRHD using 15 rules using Name, SSN, DoB, Gender, and Address:
- Last Name + (1st initial of) First Name + Gender + DoB
- Last Name (soundex) + First Name (soundex) + Gender + DoB
- Last Name + First Name + DoB + zip3
- Last Name + First Name + Gender + DoB
- SSN + Gender + DoB
- Last Name + (1st 3 characters of) First Name + Gender + DoB + zip3
- Last Name + (1st 3 characters of) First Name + Gender + DoB
- Last Name + (1st 3 characters of) First Name + Gender + zip5
- Last Name + address
- Last Name + (1st 3 characters of) First Name + Gender + zip3
- Last Name + First Name + Gender + zip3
- Last Name + First Name + Gender + zip5
- Last Name + (1st initial of) Middle Name + First Name + Gender +zip3
- Address + zip5 + DoB
- Address + zip5 + DoB + Gender

Datavant software application leaves the choice of favoring one rule over other to the research teams. A site submitting patient data to PCORNet, will attempt to submit hashes against all rules for each patient, *i.e.*, 15 per patient, prior to sending the data to PCORNet. If the underlying data element is invalid or missing, the resulting hash can be left empty and in the process, the end user, PCORNet researcher, is made aware that the particular rule is not suited for linking.

NDI first provides a probabilistic score to the match [NHIS2009]. The score is based on probabilistic weights assigned to each of the identifying data items used in the linkage of the National Center for Health Statistics (NCHS) Survey Data to the NDI [NCHS2022]. NCHS has since developed new binit weights [NCHS2009], based upon the frequency of occurrence of the data items in the NDI files for the years 1979 to 2000, which represents about 49 million persons. The weights correspond to [$\log_2 (1/p_i)$]: the base 2 logarithm of the inverse of the probability of occurrence of the value of the identifying data item on the submission record. For example, a common first name, such as "John", that has a higher probability of occurrence in the population has a lower weight than an uncommon name such as "Jonas". NDI further categorizes its matches [NHIS2009] into five mutually exclusive classes depending on the degree of agreement of the underlying data elements. The highest specificity category is class 1 and is designated to matches that agree on at least 8 (of 9) digits of SSN, first name, middle initial (including blank), last name, birth year (+/- 3 years), birth month, sex, and state of birth. In the middle, class 3 is designated to matches where SSN is unknown but eight or more of first name, middle initial, last name, father's surname (for females), birth day, birth month, birth year, sex, race, marital status, or state of birth match. The lowest is class 5, where SSN is present but fewer than 7 (of 9) digits on SSN agree. In the final presentation, it selects the one with the smallest value of class for the patient and then selects the one with the largest score. In case of a tie, a manual review is performed.

In our analysis, we will perform data standardization [Kho2015, Maaroufi2018] but do not treat other human-generated errors such as alphabet/digit transpositions. We will include the derivative Soundex for completion [Kho2015, Levin2019, Datavant] but Soundex is designed for English names. In California Bay Area, nearly 50% of the population is non-white (https://www.census.gov/quickfacts/sanfranciscocountycalifornia), and of those who identify themselves as white, a significant fraction is not expected to have English names e.g., Italian, German and French.  There are other approaches like Metaphone [Philips1990] and double Metaphone [Philips2000] which significantly improve upon the Soundex algorithm that we plan to consider for our final product.

Our final patient-matching product is designed similarly to DCIFIRHD [Kho2015] in that we can add more rules as we find new data sources to join against hospital patient data. One of the advantages of this approach is that privacy-preserving transformations (aka hashing) can be added to the product if and when needed.

# Methodology:
The underlying software framework needs to support the creation of new rules (step 3 below) starting from specific elements. For two datasets that are linked, the analysis framework needs to evaluate the quality of linking (steps 4 and 5).

**Assumptions:**
We will assume the NTIS LADMF death data to be accurate when available. However, the absence of a particular person on the LADMF file is not proof that the individual is alive.  We will also assume that EHR death data, when available, is accurate.

We will use the following elements from the LADMF dataset for linking: (First Name, Middle Name, Last Name or Surname), date of birth (DoB), date of death (DoD), and Social Security Number (SSN) if available. We ignore the Name Suffix (Alphanumeric, like SR JR, III) in the LADMF data. No other elements are available. We will use the following elements from the EHR dataset for linking: (First Name, Middle Name, Last Name or Surname), DoB, DoD, and SSN.

EHRs often present previous names, preferred names, and more. Our generalized framework supports using alternate names, but for this analysis, we limit it to using the latest name in the EHR system.

Finally, these data are human-reported/entered data and contain human error such as swapping e.g. Last and First Names, two adjacent letters, MM and DD in a date format (MM/DD/YYYY). For the analysis, we check data validity but do not attempt to correct errors.

**Step 1a: Access and pre-processing of LADMF dataset:**
There are several regulatory processes to go through to access the LADMF dataset for Stanford University research use. Upon completion of the requirements, the dataset can be downloaded in our secure Google Cloud Platform (GCP) data center [Datta2020] using sftp protocol. After an initial full download, the data is downloaded monthly. Each new dataset is converted from its raw form to a structured database, specifically a GCP BigQuery dataset. Newer information for the person replaces older information. After this pre-processing, the dataset, STARR-LADMF, is ready for linking with other Stanford datasets in STARR that meet data use agreements.

We use STARR-LADMF data from Sep 27 2022 for analysis. There is nothing specific about this date.

**Step 1b: Access and pre-processing of SHC patient dataset:**
SHC EHR vendor is Epic and the EHR data flows into the hospital's operational Clarity. A copy of the operational Clarity data is uploaded daily to our secure GCP data center [Datta2020] and converted to a GCP BigQuery dataset. This dataset is then filtered to remove certain operational data that are not accessible for secondary use cases like research. The resulting filtered dataset becomes part of Stanford School of Medicine (SoM) STARR. The STARR-SHCClarity dataset is ready for linking with LADMF or another dataset. Note that the STARR-SHCClarity is a PHI dataset and contains typical patient identifiers.

**Step 2: Cleaning of the two datasets:**
In this step, we clean and standardize the two datasets, STARR-LADMF and STARR-SHCClarity, identically. The raw elements to standardize are (First Name, Middle Name, Last Name), DoB, DoD and SSN. The volume of our data poses specific choices on the data analysis software stack. We use Google Cloud Platform BigQuery SQL for data manipulation and transformation, IPython Magics for BigQuery library to facilitate readability and execution of the code, python 3.9.5 with matplotlib and pandas libraries for data visualization, and Jupyter Notebooks to collate data processing in one notebook. The specific code for normalization and tokenization can be found in Research IT github (DD_Standarization_and_Tokenization.ipynb in https://github.com/susom/Death-data-integration). Here is a summary of the transformation:

1. Remove unwanted characters:
    a. Keep only digits 0-9 for SSN and dates and remove characters like hyphens and slashes. For example, SSN 123-35-4789 becomes 123354789 and the date 2008/12/25 becomes 20081225.
    b. Keep only alphabets in the Names and remove digits, white spaces, and special characters such as periods, commas, accents, diacritics, ampersands, hyphens, slashes, apostrophes, asterisks, etc.
2. Truncate Strings:
    a. Limit the (First Name) to <= 15 characters

   b. Limit the (Middle Name) to <= 15 characters
   c. Limit the (Last Name) to <= 20 characters
3. Uppercase (capitalization)
4. Validate variables:
   a. Exclude years before 1850 and after 2022
   b. Exclude invalid SSN. For example, the first three digits of a valid SSN will not have "000," "666," or numbers in the 900 series. The second two-digit group will not be "00", and the third four-digit group will not be "0000." Other default patterns such as 999-99-9999, 888-88-8888, 123-45-6789, 012-34-5678 are also invalid SSNs.

A variable is considered valid when the result of the variable validation process does not return a Null value or an empty string. Standardized variables from First Name, Middle Name, Last Name, DoB, and SSN are respectively named first_name, middle_name, last_name, birth_date, and ssn.

**Step 3: Definition of rules:**

We define a rule (aka token) as a derived element generated by concatenating one or more of the standardized raw elements. We use the following standardized raw elements to create tokens: first_name, middle_name, last_name, birth_date, and ssn. We can create multiple tokens from these few raw elements.

For example, ssn + birth_date is a token (token 3 in Table 1) representing the concatenation of ssn and birth_date and is derived from standardized raw elements SSN and DoB. For example, 123354789 + 20081225 from step 2 two becomes 12335478920081225 starting from raw elements 123-35-4789 and 2008-12-25. Similarly, last_name + first_name + birth_date is another token (token 11 in Table 1). **A token is generated if and only if all the standardized elements are valid.** For example, for a particular ssn+birth_date, if the ssn is null because it is either SSN is invalid or missing then the resulting token ssn+birth_date is also null. For soundex, we use Google Cloud BigQuery strong function soundex (https://cloud.google.com/bigquery/docs/reference/standard-sql/string_functions#soundex)

Table 1 below presents the tokens that were generated for the analysis. In Table 1, YYYY of birth_date is 2008 when birth_date is 20081225.

| id | Token |
|---|---|
| 1 | ssn + last_name + middle_name + first_name + birth_date |
| 2 | ssn + last_name + first_name + birth_date |
| 3 | ssn + birth_date |
| 4 | ssn + YYYY of birth_date + first_name + last_name |
| 5 | ssn + last_name + middle_name + first_name |
| 6 | ssn |
| 7 | ssn (last 4) + last_name + middle_name + first_name + birth_date |
| 8 | ssn (last 4) + birth_date |

| 9 | last_name + middle_name + first_name + birth_date |
|---|---|
| 10 | last_name + middle_name + first_name + YYYY of birth_date |
| 11 | last_name + first name + birth_date |
| 12 | last_name + 1st initial of middle_name + first_name |
| 13 | last_name + 1st 3 characters of first_name + birth_date |
| 14 | last_name + 1st initial of first_name + birth_date |
| 15 | last_name (soundex) + middle_name (soundex) + first_name (soundex) + birth_date |
| 16 | last_name (soundex) + middle_name (soundex) + first_name (soundex) + YYYY of birth_date |
| 17 | last_name (soundex) + first_name (soundex) + birth_date |
| 18 | last_name |
| 19 | first_name |
| 20 | birth_date |

Table 1: Tokens generated from standardized data elements in step 2: first_name, middle_name, last_name, birth_date and ssn.

**Step 4: Analysis steps:**
For each of the tokens defined in step 4, we run the following analysis in the two datasets.
- Complete: At what rate is this element (aka token) captured appropriately in the record? For example, the data is incomplete if the element value is null.
- Distinctive: Is this trait distinctive enough to be used as a key identifier? It is also the same as the uniqueness of the value.
- Invalid: Is this element likely to be incorrect?

For example, in the illustration below with five persons, we show the impact of these definitions:

| First Name | Date Of Birth |
|---|---|
| Jhon | Null |
| Arthur | 05/07/1950 |
| Anna | 05/07/1950 |
| %^3 | 08/08/1997 |
| Jhon | 02/03/1990 |

First Name:
- Complete=5
- Distinct=4
- Invalid=1

Date Of Birth:
- Complete=4
- Distinct=3
- Invalid=0

Table 2: An illustrated list of patient first names and date of births.

**Step 5: Validation:**

We do not have a golden dataset (aka truth set) to match our results against. Therefore, in our study, we will use algorithm match and non-match [Sequoia2018] to analyze cross-organizational patient identity:

|  | Algorithm Match | Algorithm Non-match |
|---|---|---|
| Actual Match | True Positive (TP) = Refers to the correct classification by the matching algorithm of two patient records as a match when both records refer to the same person. | False Negative (FN) = Type II error. A classification error by the matching algorithm where the two patient records are marked as referring to two distinct patients but in reality, the two records refer to the same person |
| Actual Non-Match | False Positive (FP) = Type I error. Refers to a classification error in the matching algorithm where a record pair is marked as a match but in reality, the two records refer to two distinct patients | True Negative (TN) = Refers to the correct classification by the matching algorithm of two patient records as a non-match when the two records refer to two different patients |

For each SHC patient where we have a death_date available, we define the following for each SHC token [Sequoia2018]:

- Algorithm Match: The token results in a unique reported death_date in the LADMF token database and the death_date in SHC matches the unique death_date in the LADMF database. This can come from both True Positives and False Positives. We expect this to represent True Positive if the underlying data elements are sufficiently distinct in the two datasets.
- Algorithm Non-match: The token results in a unique death_date in the LADMF token database but the death_date in SHC does not match the reported death_date in the LADMF database. The difference can be due to False Negatives and True Negatives. We expect this to represent True Negatives if the underlying data elements are sufficiently distinct in the two datasets.

## Results:

In this section, we present the quality of standardized data elements and derived tokens using the definitions - complete value, distinct values, and invalid values.

The following table, Table 3a, presents the following metrics for the standardized data elements - complete, distinctive, invalid - in the LADMF dataset. For the particular dataset (Sep 27 2022), LADMF dataset presented 102,889,867 persons. Note that while ssn is both complete and distinct, the other data elements present incompleteness as well as indistinctness. Middle name is frequently incomplete/null (53%). Overall, there are a few invalid elements in the dataset.

| Field name | Complete | Distinct | Invalid |
|---|---|---|---|
| ssn | 102,889,867 | 102,889,867 | 0 |
| last_name | 102,889,867 | 2,150,557 | 1 |
| first_name | 102,889,529 | 870,043 | 3 |
| middle_name | 102,857,738 | 249,423 | 631 |
| death_date | 102,889,866 | 32,601 | 0 |
| birth_date | 102,076,278 | 65,520 | 4,555 |

Table 3a: For 102,889,867 persons in LADMF, we present the metrics for complete, distinct, null and invalid values.

The following table, Table 3b, presents the following metrics for the standardized data elements - complete, distinctive, invalid - in the SHC Clarity dataset. For the particular dataset (Sep 27 2022), Clarity dataset presented 4,587,669 records. Overall, ssn is found to be invalid frequently (36%), or null (16%), and the death date is often null (99%).

| Field name | Complete | Distinct | Invalid |
|---|---|---|---|
| ssn | 3,866,460 | 2,201,054 | 1,668,310 |
| last_name | 4,587,660 | 710,832 | 2 |
| first_name | 4,587,651 | 350,742 | 2 |
| middle_name | 1,385,138 | 74,773 | 40 |
| death_date | 40,024 | 19,090 | 0 |
| birth_date | 4,587,170 | 52,814 | 37 |

Table 3b: For 4,587,669 records in SHC Clarity, we present the metrics for complete, distinct, and invalid values.

We further look into the ssns in SHC Clarity data. In Table 3c, for the 1,668,310 invalid entries, we find that 99% come from "default" values like 999-99-9999.

| Invalid-ssn | Count |
|---|---|
| 999-99-9999 | 921,620 |
| 888-88-8888 | 630,279 |
| 000-00-0000 | 110,365 |
| 9xx-xxx-xxx | 3,567 |
| xxx-00-xxxx | 1,908 |
| 666-xxx-xxx | 296 |
| 000-xxx-xxx | 218 |
| 001-01-0001 | 16 |
| 111-11-1111 | 13 |
| 123-45-6789/012-34-5678 | 10 |
| 090-90-9090 | 5 |
| 444-44-4444 | 4 |
| 555-55-5555 | 4 |
| 777-77-7777 | 3 |
| 666-66-6666 | 1 |
| 333-33-3333 | 1 |

Table 3c: Breakdown of invalid ssns in SHC Clarity data

While SSA had issued over 450 million original SSNs by 2008 [Puckett2009], we believe that EHR and claims datasets will continue to present significantly missing SSNs. In our pediatric Clarity, the frequency of valid SSNs is even lower, only 15% of 1.5 million patients have a valid SSN.  There are several reasons for an invalid SSN. Firstly, SSNs are required only for working individuals. Secondly, elderly patients and spouses of legal immigrants may not have an SSN. Thirdly, pediatric patients may not have an SSN at the time of their hospital stay. Fourthly, some patients choose not to disclose their SSN due to privacy concerns. There is also a growing trend to not require SSNs  in EHR and claims systems. As of April 2019, the Centers for Medicare & Medicaid Services (CMS) has removed SSNs from all Medicare cards, as required by the Medicare Access and CHIP Reauthorization Act of 2015 [MACRA2015]. The SSN was successfully removed from Mayo Clinic's EHR in April 2022.

In Tables 4a and 4b, we present the completeness and distinctiveness of the derived elements, the tokens corresponding to tokens in table 1. Tokens are only generated if the underlying element is not null and is valid. Table 4a presents the tokens for LADMF. Table 4b presents the tokens for SHC Clarity.

| id | Token | Completeness | | Distinctiveness | |
|---|---|---|---|---|---|
| | | Total records | % of total records | Total records | %of total records |
| 1 | ssn + last_name + middle_name + first_name + birth_date | 47,990,976 | 46.6% | 47,990,976 | 46.6% |
| 2 | ssn + last_name + first_name + birth_date | 102,071,443 | 99.2% | 102,071,443 | 99.2% |
| 3 | ssn + birth_date | 102,071,720 | 99.2% | 102,071,720 | 99.2% |
| 4 | ssn + YYYY of birth_date + first_name + last_name | 102,071,443 | 99.2% | 102,071,443 | 99.2% |
| 5 | ssn + last_name + middle_name + first_name | 48,425,839 | 47.1% | 48,425,839 | 47.1% |
| 6 | ssn | 102,889,867 | 100% | 102,889,867 | 100% |
| 7 | ssn (last 4) + last_name + | 47,990,976 | 46.6% | 47,990,960 | 46.6% |

| | middle_name + first_name + birth_date | | | | |
|---|---|---|---|---|---|
| 8 | ssn (last 4) + birth_date | 102,071,720 | 99.2% | 84,211,971 | 81.8% |
| 9 | last_name + middle_name + first_name + birth_date | 47,990,976 | 46.6% | 47,951,556 | 46.6% |
| 10 | last_name + middle_name + first_name + YYYY of birth_date | 47,990,976 | 46.6% | 45,383,492 | 44.1% |
| 11 | last_name + first name + birth_date | 102,071,443 | 99.2% | 101,367,840 | 98.5% |
| 12 | last_name + 1st initial of middle_name + first_name | 48,425,839 | 47.1% | 31,452,009 | 30.6% |
| 13 | last_name + 1st 3 characters of first_name + birth_date | 102,071,443 | 99.2% | 100,848,613 | 98.0% |
| 14 | last_name + 1st initial of first_name + birth_date | 102,071,443 | 99.2% | 96,449,374 | 93.7% |
| 15 | last_name (soundex) + middle_name (soundex) + first_name (soundex) + birth_date | 47,990,976 | 46.6% | 47,889,941 | 46.5% |
| 16 | last_name (soundex) + middle_name (soundex) + first_name (soundex) + YYYY of birth_date | 47,990,976 | 46.6% | 37,903,824 | 36.8% |
| 17 | last_name (soundex) + first_name (soundex) + birth_date | 102,071,443 | 99.2% | 98,600,135 | 95.8% |
| 18 | last_name | 102,889,864 | 100% | 2,080,005 | 2.0% |
| 19 | first_name | 102,889,522 | 100% | 839,943 | 0.8% |
| 20 | birth_date | 102,071,720 | 99.2% | 61,824 | 0.1% |

Table 4a: Token completeness and distinctiveness for 102,889,867 persons in LADMF dataset.

| | | Completeness | | Distinctiveness | |
|---|---|---|---|---|---|
| id | Token | Total records | % of total records | Total records | %of total records |
| 1 | ssn + last_name + middle_name + first_name + birth_date | 826,435 | 18.01% | 826,424 | 18% |
| 2 | ssn + last_name + first_name + birth_date | 2,198,483 | 47.92% | 2,198,413 | 48% |
| 3 | ssn + birth_date | 2,198,489 | 47.92% | 2,197,904 | 48% |
| 4 | ssn + YYYY of birth_date + first_name + last_name | 2,198,483 | 47.92% | 2,198,367 | 48% |
| 5 | ssn + last_name + middle_name + first_name | 826,435 | 18.01% | 826,403 | 18% |
| 6 | ssn | 2,198,506 | 47.92% | 2,195,146 | 48% |
| 7 | ssn (last 4) + last_name + middle_name + first_name + birth_date | 826,435 | 18.01% | 826,419 | 18% |
| 8 | ssn (last 4) + birth_date | 2,198,489 | 47.92% | 2,189,157 | 48% |
| 9 | last_name + middle_name + first_name + birth_date | 1,384,971 | 30.19% | 1,384,199 | 30% |
| 10 | last_name + middle_name + first_name + YYYY of birth_date | 1,384,971 | 30.19% | 1,382,483 | 30% |
| 11 | last_name + first name + birth_date | 4,587,099 | 99.99% | 4,573,575 | 99.7% |
| 12 | last_name + 1st initial of middle_name + first_name | 1,384,989 | 30.19% | 1,309,624 | 29% |
| 13 | last_name + 1st 3 characters of first_name + birth_date | 4,587,099 | 99.99% | 4,566,012 | 99.5% |
| 14 | last_name + 1st initial of first_name + birth_date | 4,587,099 | 99.99% | 4,549,792 | 99% |
| 15 | last_name (soundex) + middle_name (soundex) + first_name (soundex) + birth_date | 1,384,971 | 30.19% | 1,383,449 | 30% |
| 16 | last_name (soundex) + middle_name (soundex) + first_name (soundex) + YYYY of birth_date | 1,384,971 | 30.19% | 1,373,064 | 30% |
| 17 | last_name (soundex) + first_name (soundex) + birth_date | 4,587,099 | 99.99% | 4,560,424 | 99% |
| 18 | last_name | 4,587,658 | 99.9998% | 545,103 | 12% |
| 19 | first_name | 4,587,644 | 99.999% | 277,757 | 6% |
| 20 | birth_date | 4,587,133 | 99.99% | 46,674 | 1% |

Table 4b: Token completeness and distinctiveness for 4,587,669 records in SHC Clarity dataset.

In Table 5, we present the validation data for the 40,024 SHC patients (aka the validation subset) for whom we have death_date in EHR. We only count the tokens if they are unique in 40,024 SHC token database and have a single match in NTIS LADMF token database. If the token appears more than once in the validation subset we ignore the token. If they are unique in validation dataset but not in LADMF, we ignore them. For the tokens that present a one-to-one match, we then consider whether the death date matches or not.

| id | Token (tested against 40,024 known records ) | 1-to-1 token match | DoD match | DoD non-match | DoD match Rate |
|---|---|---|---|---|---|
| 7 | ssn (last 4) + last_name + middle_name + first_name + birth_date | 1,081 | 920 | 161 | 85.1% |
| 1 | ssn + last_name + middle_name + first_name + birth_date | 1,074 | 914 | 160 | 85.1% |
| 5 | ssn + last_name + middle_name + first_name | 1,103 | 938 | 165 | 85.0% |
| 9 | last_name + middle_name + first_name + birth_date | 1,170 | 986 | 184 | 84.3% |
| 15 | last_name (soundex) + middle_name (soundex) + first_name (soundex) + birth_date | 1,260 | 1,039 | 221 | 82.5% |
| 2 | ssn + last_name + first_name + birth_date | 3,232 | 2,647 | 585 | 81.9% |
| 4 | ssn + YYYY of birth_date + first_name + last_name | 3,296 | 2,696 | 600 | 81.8% |
| 3 | ssn + birth_date | 4,084 | 3,292 | 792 | 80.6% |
| 11 | last_name + first name + birth_date | 3,847 | 2,972 | 875 | 77.3% |
| 6 | ssn | 4,445 | 3,398 | 1,047 | 76.4% |
| 13 | last_name + 1st 3 characters of first_name + birth_date | 4,240 | 3,206 | 1,034 | 75.6% |
| 17 | last_name (soundex) + first_name (soundex) + birth_date | 4,821 | 3,098 | 1,723 | 64.3% |
| 14 | last_name + 1st initial of first_name + birth_date | 5,198 | 3,249 | 1,949 | 62.5% |
| 10 | last_name + middle_name + first_name + YYYY of birth_date | 1,572 | 934 | 638 | 59.4% |
| 8 | ssn (last 4) + birth_date | 7,611 | 2,636 | 4,975 | 34.6% |
| 12 | last_name + 1st initial of middle_name + first_name | 2,325 | 767 | 1,558 | 33.0% |
| 16 | last_name (soundex) + middle_name (soundex) + first_name (soundex) + YYYY of birth_date | 2,592 | 760 | 1,832 | 29.3% |
| 19 | first_name | 427 | 56 | 371 | 13.1% |
| 18 | last_name | 521 | 66 | 455 | 12.7% |
| 20 | birth_date | 79 | 4 | 75 | 5.1% |

Table 5: Number of matches and non-matches for the validation dataset of 40,024 SHC patients for whom we have death_date in EHR. The table is ordered by match rate.

## Discussion:

Based on the results in Table 5, a good linking approach is to use information where available. For example, token 1 (ssn + last_name + middle_name + first_name + birth_date) has the most information and therefore has the least error, but is frequently incomplete.

There are some additional death date validations that may be implemented. For example, if the patient presents certain types of EHR encounters beyond the linked death date, then that death date is bogus. If the EHR data is linked to another EHR or claims dataset (here, a higher quality probabilistic model can be developed), and if there is an encounter or claim after a linked death date, then the death date is likely bogus. A model imputation can add another dimension thereby increasing the utility of data. In our case, when a patient is no longer observed in EHR, it could be because they discontinued/switched hospital or because they are deceased. Reps *et al* [Reps2019] use the SSA DMF augmented claims dataset for Discriminating End of observation into Alive and Dead (DEAD) model development and validate the model on three other claims dataset. All datasets are in OMOP format and all the code required to apply the developed model on new data is available on github (https://github.com/OHDSI/StudyProtocolSandbox/tree/master/DeadModel). The model variables used are available in a typical EHR dataset. Kim *et al* [Kim2021] create another model to predict the cause of death using South Korea national claims and EHR data and validate it on US claims datasets. Here again, the model development uses OMOP and the code is publicly available on github (https://github.com/ABMI/CauseSpecificMortality). Furthermore, it may be possible to generate probabilistic weights similar to NDI weights [NCHS2022] or NCHS binit weights [NCHS2009] or reuse NCHS binit weights if they are made publicly accessible.

At this time, the r-CDWs in STARR do not support death date validation by checking encounter types, we do not compute probabilistic weights or implement death models. Our ultimate goal is to include other sources of death data. For example, ~90% of SHC patients reside in the state of California and the California non-comprehensive Fact of Death (FOD) file is a good candidate for addition. We also wish to incorporate machine learned models such as DEAD [Reps2019]. For our final set of tokens, we plan to use historical and preferred names available in EHR, and use phonetic matching of names such as double metaphones. At this time, the r-CDWs do not support these alternate names either.

In the context of a r-CDW, the presentation of cross-organizational linked data exists to support myriad different use cases. How does a r-CDW share this uncertainty with the end user, in the context of FAIR guidelines of data stewardship? Linkage errors can bias the results of data analyses unless linkage method details are reported in detail, or algorithms are validated against good-quality reference standard data sets. We believe that the following information will help the end users make better use of the data:

1. Actual code for raw data standardization, e.g., DD_Standarization_and_Tokenization.ipynb in https://github.com/susom/Death-data-integration. This information allows the user to understand how the rules are generated.
2. For every two datasets that are linked, an analysis of completeness and distinctiveness of tokens, e.g., Tables 4a and 4b. This information allows the user to understand the potential missingness and noisiness inherent in the source data prior to linking.
3. An analysis of the algorithmic match and no-match, *e.g.*, Table 5. In our case, we are able to match on a small fraction of total patients (~40,000 in 4 Million). This information allows the user to assess the trustworthiness of specific token linkages for their use case.

Finally, for patient-level presentation of any two linked datasets, we present only those tokens that are unique in the two datasets i.e., 1-to-1 match in the order of priority of higher specificity tokens first. The 1-to-1 match is a simplistic mechanism of retaining top scoring data elements (most unique elements gets the top score and only the top scoring candidate is retained). One challenge of this approach is that as the underlying data changes, the uniqueness of tokens will diminish in time and will result in death date instability between refreshes.

We also classify the tokens in the following categories:
- Category 1: Token ids in validation set that have >80% DoD match rate *i.e.*, tokens 7, 1, 5, 9, 15, 2, 4, 3
- Category 2: Token ids in validation set that have 50-80% DoD match rate *i.e.*, tokens 11, 6, 13, 17, 14, 10
- Category 3: Token ids in validation set that have <50% DoD match rate *i.e.*, tokens 8, 12, 16, 19, 18, 20

See Table 7 for an illustration. Keeping the class definition constant, we can add more tokens and more data sources. Within each data source, we simply present the DoD corresponding to the superior token first. For example, in class 1, 7 is better than 15, so if a DoD is available for both 7 and 15, only 7 will be presented.

| Patient ID | DoD in SHC | DoD in LADMF \| Category \| Token id |
|---|---|---|
| 1001 | null | 7/17/1993 \| 1 \| 1 |
| 1002 | 12/3/2007 | 12/7/2007 \| 2 \| 6 |
| 1003 | null | 4/5/2021 \| 3 \| 8 |

Table 7: Illustration of how deterministically linked death data from LADMF is included in a r-CDW.

Armed with the above information, the users can make use of the linkage details depending on their specific use case. For example, in the case of readmission models [Liu2021], the models can present their analysis using no death information, adding EHR death information only, further adding in the death information from category 1 (or 2) of LADMF and observing how the models are impacted as one or other dates are used. In another example, the end-user may use hybrid approaches such as those used by Eisenstein *et al* [Eisenstein2019] where the research team takes some information from the LADMF and use other resources such as obituary data or NDI, thereby limiting effort or cost.

## Acknowledgement:

Using CRediT taxonomy (https://credit.niso.org/), we present the contributing roles for manuscript authors - Alvaro Alvarez (Methodology, Software, Formal Analysis, Writing - review and editing), Priya Desai (Investigation, Resources, Supervision), Somalee Datta (Conceptualization, Visualization, Writing - original draft). David Love, and Joseph Pallas from Research IT team support NTIS LADMF data acquisition and pre-processing (step 1a) respectively. Deepa Balraj and Joe Mesterhazy support pre-processing (step 1b) of SHC patient dataset. Several other team members provided technical feedback and support including Tina Seto, Nitin Parikh, Andrew Martin, Jae Lee and De Lin (Darren) Guan.